\documentclass[amsmath,showpacs,twocolumn,aps,prl]{revtex4}

\usepackage{amsmath,amsfonts}
\usepackage{graphicx}
\usepackage{color}
\usepackage{enumerate}

\date{\today}

\begin{document}

\title{Majorana edge states in interacting one-dimensional systems}

\author{Suhas Gangadharaiah,$^1$ Bernd Braunecker,$^1$ Pascal Simon,$^2$ and Daniel Loss$^1$}
\affiliation{$^1$Department of Physics, University of Basel,
             Klingelbergstrasse 82, 4056 Basel, Switzerland\\
             $^2$Laboratoire de Physique des Solides, CNRS UMR-8502,
             Univ. Paris Sud, 91405 Orsay Cedex, France}


\begin{abstract}
We show that one-dimensional  electron systems in proximity of a
superconductor that support Majorana edge states are extremely
susceptible to electron-electron interactions. Strong interactions generically
destroy the induced superconducting gap that stabilizes the Majorana
edge states. For weak interactions, the renormalization of the gap is
nonuniversal and allows for a regime, in which the Majorana edge states
persist. We present strategies how this regime can be reached.
\end{abstract}

\maketitle

\emph{Introduction.}
The possibility of realizing Majorana bound states at the ends of one-dimensional (1D)
conductors formed by topological
insulator edge states, semiconductor nanowires or carbon nanotubes
in the proximity of a superconductor
\cite{fu_kane,alicea,sau,sau1,sau2,oreg,alicea2,klinovaja},
as well as by quasi-one-dimensional superconductors \cite{potter_lee}
has led recently to much activity.
An important factor for the interest is the potential application of the Majorana
edge states as elementary components of a topological quantum computer \cite{kitaev,Hassler,TQC,alicea2}.
In a nanowire the Majorana edge modes exist because of the  $p$-wave nature of the induced superconductivity,
which is the result of the projection of the superconducting order parameter onto the
band structure of the wire, consisting of helical, i.e., spin (or Kramers doublet)
filtered left and right moving conducting modes.
In such a setup, the Majorana edge states
appear as particle-hole symmetric Andreev bound states at both ends of the wire, with a localization length $\xi$
inversely proportional to the induced superconducting gap $\Delta$, and their
wave function overlap is typically proportional to $\exp(- L / \xi)$ with $L$ the wire length.
The independence and the particle-hole symmetry
of the two bound states is only guaranteed if this
overlap is vanishingly small, therefore large $L$ and $\Delta$ are required.

Electron-electron interactions renormalize the properties of one-dimensional
conductor and so modify $\Delta$ as well as the localization length of the  bound states.
In this paper, we study these interaction effects in systems
with helical conduction states that are in contact with a superconductor.
We show that superconductivity
and Majorana edge states are stable only at weak interactions.
Strong and long-ranged interactions generically suppress superconductivity and so delocalize
and suppress the Majorana edge states. For weaker and screened interactions,
superconductivity and the Majorana edge states remain stable only if the renormalization flow reaches
the strong coupling limit for the induced superconducting gap at a correlation length
$\xi \ll L$. This regime is reached for a large induced superconducting
gap, best possible screened interactions, and the longest possible wire length $L$,
which outlines the necessary strategy  in the experimental search for Majorana edge states.
Under these conditions, although the electron interactions in most cases substantially
reduce the size of the gap, the Majorana edge states remain strongly localized
at each end.

In the following, we first illustrate the effect of electron interactions
on the Majorana bound states using the fermion chain model of
Ref. \cite{kitaev}. In particular, we show that for strong interactions
the gap can entirely close and the system becomes equivalent to a gapless free electron gas.
Motivated by this insight, we turn to a continuum theory for the nanowires,
allowing us to include the interactions more effectively and to move beyond
the restriction to a half-filled chain.


\emph{Fermionic chain.}
The prototype model for Majorana edge states is a one-dimensional  open
lattice of sites $i=1,\dots,N$ described by the model \cite{kitaev,kogut}
\begin{equation}
	H =
	-
	\sum_{i=1}^{N-1}
	\left[
		t c_i^\dagger c_{i+1} + \Delta c_i^\dagger c_{i+1}^\dagger
		+
		\text{h.c.}
	\right]
	-
	\mu \sum_{i=1}^N n_i,
\end{equation}
where $c_i$ are tight-binding operators of spinless fermions,
for example  the electron operators of the helical conduction bands,
$t>0$ is the hopping integral,
$\Delta>0$ the triplet superconducting gap,
$\mu$ the chemical potential,
and $n_i = c_i^\dagger c_i$.
In terms of the Majorana fermion basis~\cite{Majorana}
$\gamma_i^1 = c_i + c_i^\dagger$ and
$\gamma_i^2 = i (c_i - c_i^\dagger)$,
the model is rewritten as
$H =
	-
	i \sum_{i=1}^{N-1}
	\left[
		w_+ \gamma_i^2 \gamma_{i+1}^1
		-
		w_- \gamma_i^1 \gamma_{i+1}^2
	\right]
	-
	i \frac{\mu}{2}
	\sum_{i=1}^N \gamma_i^2 \gamma_i^1$,
with $w_\pm = (t\pm \Delta)/2$.
At $t = \Delta$ and $\mu=0$, the only nonzero interaction is $w_+$, and the
ground state corresponds to pairing of Majorana fermions between neighboring
sites $\gamma^2_i \gamma^1_{i+1}$, with an excitation gap of $2w_+$.
In the open chain, $\gamma^1_1$ and $\gamma^2_N$  no longer appear in
$H$ and remain unpaired.
They form the two Majorana bound states that are localized on a single lattice site
at each edge of the wire and can be occupied at no energy cost.
For $\mu \neq 0$ or $\Delta \neq t$, the two edge Majorana modes
are coupled to the bulk system and their spatial extension becomes larger,
on the order of $\xi \sim a/\ln|w_+/w|$, with $w = \max\{|\mu|,|w_-|\}$
and $a$ the lattice constant.
In the finite system, the overlap of the two Majorana states at both ends of
the chain is proportional to $e^{-N a/\xi}$, and the two states are independent
only for $Na \gg \xi$.

In such a system, interactions between the fermions critically affect
the existence and stability of the Majorana edge states. Indeed, they
lead not only to a further coupling of the Majorana edge states to the bulk system,
but also can substantially reduce the bulk gap size.
As an illustration, we include into the model the
repulsive nearest neighbor interaction
$H'	= U\sum_{i=1}^{N-1}  \left(n_i-1/2\right) \left(n_{i+1}-1/2\right)$,
with $U>0$. The mean field contribution of this interaction is inessential. The direct part
can be removed by further tuning $\mu$ as well as the edge potentials, while the exchange part adds
to $w_\pm$ and the $w_-$ contribution can be removed by tuning $t-\Delta$.
Quantum fluctuations, however, cannot be suppressed, and their role reaches much further.

Indeed, it is straightforward to show that interactions can entirely close the superconducting
gap. For strongly interacting $t=\Delta=U/4$ we can map $H$ by a Jordan-Wigner transformation to
the spin chain $H = t \sum_{i=1}^{N-1} (\sigma_i^x \sigma_{i+1}^x + \sigma_i^z \sigma_{i+1}^z)$,
where $\sigma^{x,y,z}_i$ are spin 1/2 operators (normalized to $\pm 1$) defined by
$c_i = \frac{1}{2}(\sigma_i^x + i \sigma_i^y) \prod_{j<i} \sigma_j^z$. By a further Jordan-Wigner
transformation to new fermion operators $\tilde{c}_i = \frac{1}{2} (\sigma_i^z + i \sigma_i^x) \prod_{j<i} \sigma_j^y$
we then see that $H = - 2t \sum_{i=1}^{N-1} (\tilde{c}_i^\dagger \tilde{c}_{i+1} + \tilde{c}_{i+1}^\dagger \tilde{c}_i)$,
which describes a free gapless fermion gas in which the localized states have disappeared.

The role of interactions are therefore crucial for understanding the
stability and existence of the Majorana edge states. In the following  we use a continuum description for a
quantitative analysis, which allows us to include the interactions more effectively,
first at half filling as in the discrete model, then away from half filling.

\emph{Continuum model.}
For the continuum theory, we focus on a quantum wire
with Rashba spin-orbit interaction in a magnetic field
with proximity induced singlet superconductivity \cite{sau,sau1,sau2,oreg}.
The non-interacting part of the Hamiltonian for the quantum wire can
be written as a sum of two parts, $H_0=H_0^{(1)} +H_0^{(2)}$, where $H_0^{(1)}$
is given by (throughout the paper $\hbar=1$)
\begin{equation}
	H_0^{(1)}
	= \int dr \Psi^\dagger_\alpha
	\biggl[
		\left(\frac{p^2}{2m} -\mu\right)\delta_{\alpha\beta}
		+
		\alpha_R p~ \sigma^{x}_{\alpha\beta}
		-
		\Delta_Z\sigma^{z}_{\alpha\beta}
	\biggr]
	\Psi_\beta,
\label{eq:nonI1}
\end{equation}
where
$\Psi_\alpha$ is the electron operator for spin $\alpha$,
the summation over repeated spin indices, $\alpha, \beta$, is assumed,
$r$ is the coordinate along the wire,
$p = -i \partial_r$,
$\alpha_R$ is the spin-orbit velocity, and
$\Delta_Z$ is the Zeeman energy of the magnetic field applied along the spin $z$
direction perpendicular to the spin-orbit selected spin $x$ direction.
The second part, $H_0^{(2)}$, includes the induced singlet superconducting term
with order parameter $\Delta_S$ and is expressed as, $	H_0^{(2)}
	= i\int dr
	\Delta_S\Psi^\dagger_\alpha\sigma^{y}_{\alpha\beta}\Psi^\dagger_\beta/2 +\text{ h.c.}$
Without interactions, $H_0^{(1)}$ has the eigenvalues
$\epsilon_{\pm} = p^2/2m  \pm \sqrt{(\alpha_R p)^2  + (\Delta_Z/2)^2}$
and corresponding eigenmodes $\Psi_\pm(p)$.
Expanding the singlet superconducting term in this eigenbasis leads to superconducting order parameters of the
triplet (within $\Psi_-$ and $\Psi_+$ subbands) as well as of the singlet type (mixing $\Psi_-$ and $\Psi_+$ subbands).
The Majorana edge states require triplet pairing \cite{sato,sau,sau1,sau2,alicea,lee,oreg,alicea2},
which is achieved by tuning the chemical potential to lie within the magnetic field gap
such that only the $\Psi_-$ subband is occupied.  In Ref.~\cite{oreg}, Majorana edge modes were
derived using the full Hamiltonian $H_0^{(1)}+H_0^{(2)}$ and were shown to exist
in the limit $\Delta_Z > \sqrt{\Delta_S^2+\mu^2}$.
  The same physics is also obtained by restricting to the occupied
$\Psi_-$ subband, which will be assumed in the following.
For $\Delta_Z \gg \Delta_S,~\alpha_R k_F$, with $k_F\approx \sqrt{m\Delta_Z}$,
the pairing then takes the compact form \cite{sato,sau,sau1,sau2,alicea,lee,oreg,alicea2}
 \begin{eqnarray}
H^{(2)}_0 \approx (\Delta/k_F)\int dr \Psi^\dagger_-(r) p\Psi^\dagger_-(r)+\text{h.c.},\label{eq:triplet}
\end{eqnarray}
with the effective triplet superconducting gap $\Delta =\Delta_S(\alpha_R k_F/\Delta_Z)$.

In the following  we work in the diagonal basis~\cite{sun}  with
the fermions confined in the $r>0$ region.
The open boundary condition forces the fermion fields to vanish
at the boundaries, thus $\Psi_{-}(r=0)=\Psi_{-}(r=L)=0$, where $L$ is the length of the wire.
In terms of the slowly varying right, $\mathcal{R}(r)$, and left, $\mathcal{L}(r)$,  moving fields,
the field $\Psi_{-}(r)$ acquires the form,
 $\Psi_{-}(r) = \sum_k \sin(k r) c_{-}(k) = e^{ik_Fr} \mathcal{R}(r)  + e^{-ik_Fr} \mathcal{L}(r)$, where
 $c_{-}(k)$ is the annihilation operator in the $\Psi_{-}$ subband.
We note that  $\mathcal{R}(r)=-\mathcal{L}(-r)$.
Thus, the kinetic energy can be expressed in terms of $\mathcal{R}$ alone
by
$H_0^{(1)} = -i v_F\int_{-L}^L dr \mathcal{R}^\dagger(r)\partial_r\mathcal{R}(r)$,
while the triplet-superconducting term acquires the form
$H_0^{(2)} \approx -\Delta \int_{-L}^L dr \,\text{sgn}(r)\bigl[\mathcal{R}^\dagger(r)\mathcal{R}^\dagger(-r)  + \text{h.c.}\bigr]$.
The noninteracting case can therefore be written as
$H_0= \int_{-L}^{L}dr~\mathbf{R}^\dagger(r) \mathcal{H} \mathbf{R}(r)$, with
\begin{eqnarray}
	\mathcal{H}=
	\left(\begin{array}{ccc}
		-i \frac{v_F}{2} \partial_r &   -\Delta \text{sgn}(r)\\
		-\Delta \text{sgn}(r) & i\frac{v_F}{2} \partial_r
\end{array}\right)\label{eq:MatrixH}
\end{eqnarray}
and $\mathbf{R}(r) = [\mathcal{R}(r),\mathcal{R}^\dagger(-r)]^T$.
Using $\mathbf{R}(r)= (e^{i3\pi/4}/\sqrt{2})\sum_{\epsilon} [u_\epsilon(r), v_\epsilon(r)]^T \gamma_\epsilon  $, where the normalized functions $u_\epsilon(r)$ and $v_\epsilon(r)$  satisfy the eigenvalue equation
$\mathcal{H} [u_\epsilon(r), v_\epsilon(r)]^T =\epsilon [u_\epsilon(r), v_\epsilon(r)]^T $, we obtain $H_0=\sum_\epsilon \epsilon \gamma_\epsilon^\dagger \gamma_\epsilon$. For $\epsilon=0$ there exists a localized mode at each edge.
At $r=0$ it is of the form  $ u_{\epsilon=0}(r)\propto e^{-2\Delta |r|/v_F}$,
with  $v_0(r) = i u_0(r)$. The operator corresponding to the edge mode,
$\gamma_{0} =\int dr u_0(r)\mathcal{R}(r) $, satisfies
the Majorana condition  $\gamma_{0}=\gamma_{0}^\dagger$.
Thus  the Majorana edge mode obtained by combining the right and
left modes  is given by,
\begin{equation}
\Psi_{\epsilon=0}^M(r)= C \gamma_0\sin(k_Fr)e^{- r/\xi},\label{eq:Majorana}
\end{equation}
for $L \gg \xi$, where $C$ is the normalization constant and  $\xi=v_F/2\Delta$ the
localization length. Note that in 1D the decay is purely exponential.
It is interesting to note that in exact analogy with the discrete lattice model,
out of the two possible Majorana states that can be constructed from the fermion field,
the localized Majorana that we obtain at one edge
corresponds to the choice  $\Psi + \Psi^\dagger$. Analogously, the Majorana localized at the other edge corresponds to $[\Psi-\Psi^\dagger]/i$.
Moreover, similar to the  edge modes in the discrete model, those obtained in the continuum limit vanish
at alternate sites for half-filling. However, the result obtained in the continuum limit is valid even away from half-filling and so more general.

\emph{Interaction effects.}
Next we include interactions between the fermions given by
$\int dr dr' V(r-r') \rho(r) \rho(r')$ with $V(r)$
being the repulsive potential and $\rho(r)$ the fermion density.
Interactions in general reduce $\Delta$, and as a consequence
$\xi$ increases. To analyze this effect,
we bosonize the Hamiltonian taking into
consideration that the low-energy physics is described by a single species of fermions
in the $\Psi_-$ subband. Using the standard procedure~\cite{giamarchi}, the bosonic Hamiltonian reads,
\begin{eqnarray}
	&&H= \int \frac{dr}{2}
	\Big[
		v_F K(\partial_r\theta)^2
		+
		\frac{v_F}{K}(\partial_r\phi)^2
		+
		\frac{4\Delta}{\pi a} \sin(2\sqrt{\pi}\theta)
 \nonumber\\
	&&  	
		-\frac{U}{\pi^2 a} \cos(4\sqrt{\pi}\phi -4k_F r)
	\Big],
\label{eq:bosH}
\end{eqnarray}
where $a$ is the lattice constant, the
$\partial_r\phi$ field describes the density fluctuations and the $\theta$ is the conjugated field.
The quadratic part in Eq.~(\ref{eq:bosH}) includes repulsive interaction
between the fermions $(K<1)$,
the sine term is due to the
triplet superconducting term $H_0^{(2)}$ given in Eq.~(\ref{eq:triplet}), and the cosine term describes umklapp  scattering by $V(r)$.
The umklapp terms play a role only in lattice systems but are
absent in quasi-one-dimensional quantum wires fabricated on a two-dimensional electron gas.
For fermions on a lattice near half-filling, $4(k_F-\pi/2a)L\ll 1$ and the oscillatory part inside
the cosine term can be neglected. The interactions then lead to the renormalization of the coupling
constants $\Delta$, $U$, and $K$, which by standard renormalization group (RG) theory~\cite{giamarchi} is
expressed by the RG equations
\begin{eqnarray}
	&&\frac{d\ln K}{dl} = \frac{\delta^2}{2K} - 2K y^2,\label{eq:RGK}\\
	&&\frac{d\delta}{dl} = (2-\frac{1}{K})\delta,~~\frac{dy}{dl} = (2-4K)y,
\label{eq:RGumklapp}
\end{eqnarray}
where   $l=\ln[a/a_0]$ is the flow parameter with $a_0$ the initial value of the lattice constant. The dimensionless
superconducting term at the length scale $a$ is defined as
$ \delta(l) = 4  a \Delta(l)/v_F$ and $y(l)=U(l) a/\pi v_F$. The initial
values of the rescaled parameters
are given by $K_0$, $\Delta_0$, $\delta_0$, $U_0$,
and $y_0$.
For $K < 1/2$ the umklapp term is relevant and superconductivity irrelevant, leading to a Mott phase,
whereas for $K > 1/2$ the opposite is true and the system is superconducting.
Near $K=1/2$ the low-energy physics depends critically on the relative strength of $\delta_0$ and $y_0$.
A large $\delta_0$ compared to $y_0$ favors superconductivity over the Mott phase and vice-versa.
An interesting scenario corresponds to the line of fixed points $\delta_0 = y_0$ and $K_0=1/2$,
where the parameters remain invariant under the RG flow.
Following Refs.~\cite{giamarchi,giamarchi2}, we find that under a change of quantization axis the theory
is described by a quadratic Hamiltonian. Therefore, similar to the discrete model with $t=\Delta=U/4$,
the spectrum is gapless.
The Majorana edge states are thus absent on the line of
fixed points, as well as in the Mott phase.

Away from half-filling, the umklapp term in Eq.~(\ref{eq:bosH}) becomes
 strongly oscillating and
can be neglected, allowing us to set $y = 0$ in Eq. (\ref{eq:RGumklapp}).
The remaining RG equations reduce to the standard Kosterlitz-Thouless (KT) equations under the change of
variables $K\rightarrow 1/2\bar{K}$ and $\delta\rightarrow \bar{\delta}/ \sqrt{2}$~\cite{giamarchi}.
The flow equation of $\Delta(l)$  is, $d\Delta/dl = (1-K^{-1})\Delta$, and its solution in terms of $K(l)$ is given by
\begin{eqnarray}
	\Delta(l) = \Delta_0 \frac{\sqrt{  8[K(l)-K_0] -4 \ln[ K(l)/K_0]    +\delta_0^2}}{\delta_0\exp[l]}, \label{eq:K1}
\end{eqnarray}
where for small deviations of $K$ from  its initial value $K_0$, $l$ is given by,
\begin{eqnarray}
l \approx \frac{K_0}{\sqrt{\alpha}}  \cot^{-1} \Bigg[   \frac{\alpha + k_0  (k_0 + x)}{x \sqrt{\alpha}}\Bigg],{}\label{eq:K2}
\end{eqnarray}
where $x = (K-K_0)/K_0$, $ k_0 = 2K_0-1$, and $\alpha = \delta_0^2/2-k_0^2$.
 \begin{figure}[ht]
  \centering
   \includegraphics[width=3.5in]{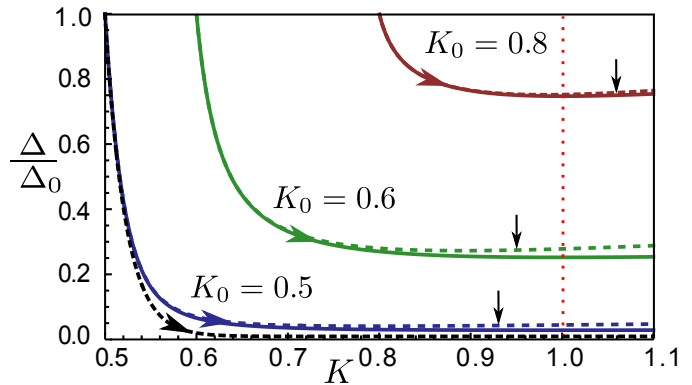}
       \caption{\label{fig:RGflow2}
	   RG flow of $\Delta/\Delta_0$ as a function of $K$ for $\Delta_0=0.05 v_F/a_0$ and the three initial values $K_0=0.5$, $K_0=0.6$, and $K_0=0.8$. The solid lines are obtained from the numerical integration of the KT eqs.  The dashed lines are obtained from  Eqs.~(\ref{eq:K1}) and~(\ref{eq:K2}) [the dashed line with the steepest decay for $K_0=0.5$ is obtained from Eq.~(\ref{eq:K1}) and $l\approx (2K_0/\delta_0^2) x$ ].
 The flow reaches the non-interacting limit at $K=1$ (shown by the red dotted line). The vertical arrows indicate the position where $\delta=1$ is reached.
	   }
\end{figure}
Rather than linearizing the KT flow eqs. around the fixed point as  is often done~\cite{giamarchi}, the solutions given by  Eqs.~(\ref{eq:K1}) and~(\ref{eq:K2}) are obtained by integrating the KT equations.
Figure~\ref{fig:RGflow2} shows  $\Delta/\Delta_0$
as a function of $K$ for $\Delta_0=0.05 v_F/a_0$ and three different values of $K_0$, $K_0=0.5,~0.6$ and $0.8$.
For all the $K_0$'s considered, $\Delta$ reduces from its initial  value and acquires its minimum at $K=1$. Note that   near $K=1$,  $\Delta$ shows very  little variation. For the strongly
repulsive case, $K_0=0.5$, $\Delta$ is reduced by  an order of magnitude as $K$ reaches $K\lesssim 1$.  In particular,  for $K\approx 0.5$ and $x\ll 1$, Eq.~(\ref{eq:K2}) can be approximated as  $l \approx  (2K_0/\delta_0^2) x $ and thus $\Delta$ has an exponential drop. More generally, the exponential decay persists as long as $x \ll \delta_0^2/ (2\max\{ k_0 ,\sqrt{|\alpha|}\})$ is satisfied. At $x\sim  \delta_0^2/ (2\max\{ k_0 ,\sqrt{|\alpha|}\} )$,  one has to consider the full form for $l$ as given by Eq.~(\ref{eq:K2}).

Next we discuss in detail the RG flow of the parameters and
its consequence for the Majorana edge states.
Although  everywhere in the repulsive regime ($K<1$) $K$ has a monotonic increase
and $\Delta$ a monotonic decrease, the flow can be divided into two
regions based on the initial values of $\delta_0$ and $K_0$.
The most favorable scenario for the existence of the Majorana edge modes corresponds to
the initial value $(K_0,\delta_0)$ with $K_0 >1/2$ in the screened regime or, for $K_0<1/2$ with  $\delta_0>2\sqrt{2K_0-\ln (2K_0e)}$ (i.e., above the separatrix). In these regions the flow is towards the strong coupling regime, and although
$\Delta$  decreases monotonically it remains finite. The minimum is reached at the  length scale
$a(l_1)$, where $K(l_1)=1$, beyond which point $\Delta$ increases. The RG flow crosses $K=1$
if the length scale $a(l_1)$ is  shorter than any cut-off length, i.e., $a(l_1)<\min\{  L,L_T,a(l_\delta) \}$ [where $l_\delta$
is defined as $\delta(l_\delta)=1$  and  $L_T=v_F/k_B T$ is the thermal length]. We note that $K=1$ is
a special line where  the interaction has scaled down to zero and the solution
is obtained exactly as in the non-interacting case without further resorting to the RG. The Majorana
edge state has the same form as in Eq.~(\ref{eq:Majorana}), albeit $\Delta$ is now given by the
reduced value $\Delta(l_1)$. However, for preserving the Majorana property, which is of particular interest for the quantum computational
use of the Majorana edge states~\cite{alicea2,kitaev}, the two edge states must have minimal overlap, i.e.,
$\chi \equiv 2\Delta(l_1) L/v_F \gg 1$.  Thus the drop in $\Delta$ due to the interactions should be compensated
 by increasing the length of the wire by at least a factor of $\Delta_0/\Delta(l_1)$, where $\Delta(l_1)$ can be evaluated from Eqs.~(\ref{eq:K1})
and~(\ref{eq:K2}). If, however, $a(l^*) = \min\{  L,L_T,a(l_\delta) \}< a(l_1)$ then the RG will be
cut-off before  $K=1$ is reached.
In the scenario when $K(l^*)\lesssim 1$, we note from  Fig.~\ref{fig:RGflow2} that $\Delta(l^*)\approx\Delta(l_1)$, thus we  expect
that the Majorana edge state will still be described by Eq.~(\ref{eq:Majorana}) with $\Delta=\Delta(l^*)$.
The second regime is the unscreened regime with $K_0<1/2$  and $\delta_0 < 2\sqrt{2K_0-\ln (2K_0e)}$.
Here the flow is towards the line of Luttinger-liquid fixed points, $\Delta=0$
 and $K_0<K<1/2$. In a realistic scenario the flow is stopped
 before the fixed points are reached at a length scale given by, $a(l^*)= \min\{L,L_T\}$. If $a(l^*)=L_T$, then $\Delta(l^*)< k_B T$
 and thermal fluctuations overcome superconductivity. On the other hand, if the wire-length $L$ is the cut-off,
 then the superconducting term is renormalized down to $\Delta(l^*) \approx \Delta_0 (L/a_0)^{1-1/K_0}$. In either
 case the bulk spectrum remain gapless and all correlations exhibit power-law decay. Thus, the Majorana edge states
 which require the presence of gapped bulk modes are absent. One way to ensure a gapped phase in the
 bulk is to consider a larger value for $\delta_0$.  A large $\delta_0$  will be difficult to achieve as the proximity induced gap $\Delta_S$ is further suppressed by the small ratio, $\alpha_R k_F/\Delta_Z$. Moreover, in contrast to $K_0$, controlling and scaling up the strength of the superconducting order parameter is non-trivial. A simpler alternative would be to apply gates on top of
the wire to screen the interactions and to increase $K_0$ to a larger $K_0'$ that
pushes the initial point $(K_0',\delta_0)$ above the separatrix,
$\delta_0>2\sqrt{2K_0'-\ln (2K_0'e)}$ or beyond $K_0'>1/2$, so that the flow is towards the strong coupling
regime.

Potential candidate systems for the observation of Majorana edge states
are the helical conductors formed at the boundaries of topological
insulators \cite{koenig,hasan_kane}, InAs nanowires with
strong spin-orbit interaction~\cite{alicea,oreg,fasth,nadj-perge}, quasi-1D unconventional superconductors~\cite{potter_lee}, carbon nanotubes
\cite{klinovaja}, and quantum wires  with nuclear spin ordering
\cite{bb2}. The latter two systems may be particularly interesting
because they are readily available and support helical modes without
external magnetic fields.

\emph{Acknowledgements.} We acknowledge discussions with C. Bourbonnais, O. Starykh, and L. Trifunovic.
This work is  supported by the Swiss NSF, NCCR Nanoscience (Basel),
and DARPA QuEST.


\end{document}